\documentclass[aps,pra,reprint]{revtex4-2}

\usepackage[english]{babel}
\usepackage[utf8]{inputenc}
\usepackage{amsmath}

\DeclareMathOperator{\arccot}{arccot}

\usepackage{amssymb}
\usepackage{graphicx}
\usepackage[colorinlistoftodos]{todonotes} 
\usepackage{hhline}
\usepackage{textgreek}
\usepackage{gensymb}
\usepackage{geometry}
\usepackage{bbold}
\usepackage{fontenc}
\usepackage{braket}
\usepackage{array}
\usepackage{subcaption}
\usepackage{soul}
\usepackage{blindtext}
\newcommand{\beq}{\begin{equation}}
\newcommand{\eeq}{\end{equation}}
\newcommand{\bea}{\begin{eqnarray}}
\newcommand{\eea}{\end{eqnarray}}

\newcommand{\tr}{\mbox{Tr}}

\begin{document}

\title{Detecting Initial System-Environment Correlations in Open Systems}
\author{Sarah Hagen}
\author{Mark Byrd}
\affiliation{Department of Physics, Southern Illinois University, Carbondale, IL 62901, USA}

\begin{abstract}
Correlations between a system and its environment lead to errors in an open quantum system.  Detecting those correlations would be valuable for avoiding and/or correcting those errors.  Here we show that we can detect correlations by only measuring the system itself if we know the cause of the interaction between the two, for example in the case of a dipole-dipole interaction.  We investigate the unitary $U$ which is associated with the exchange Hamiltonian and examine the ability to detect initial correlations between a system and its environment for various types of initial states.   
The states we select are motivated by realistic experimental conditions and we provide bounds for when we can state with certainty that there are initial system-environment correlations given experimental data.  
\end{abstract}

\maketitle


\section{Introduction}
\label{sec:introduction}

Entanglement is a uniquely quantum mechanical property and one that is responsible for many of the advantages that quantum systems have over their classical counterparts.  Its detection and manipulation are therefore of great importance in the search for the implementation of practical quantum technologies \cite{Horodeckientrev,Nielsen/Chuang:book}.  When there is entanglement in a bipartite quantum system, and one has access to both subsystems, one can perform measurements on both to detect its presence.  

However, entanglement, and more generally, correlations, can be a problem when they are unwanted.  Unwanted interactions between the system and environment cause noise in the quantum system that leads to errors in the quantum information and/or quantum control of the system.  Such errors can cause irreparable damage to the state of the system and must be avoided, suppressed, and/or corrected to achieve accurate control \cite{lidar_brun_2013}.  Like entanglement, more general correlations can be detected if both parts of a bipartite system can be accessed, controlled, and measured.  But this is often not the case for errors which are caused by an unwanted interaction with the environment.  This is because many environments are not readily accessible in the sense that they are not controllable or measurable except, perhaps, for some bulk properties.  

In the case that correlations arise between a system and environment, or some other inaccessible quantum system, what can be done to detect properties of the correlations to aid in error prevention and control?  Fairly recently, important progress has been made toward detecting correlations between a system and its environment by making measurements only on the system.  For example, if one prepares many different initial states, one can use the method proposed in \cite{Modi_2012,Modi:2015} to find whether the effective environment state depends on the preparation of the system \cite{Ringbauer:2015}.  This process requires many preparations and measurements.  Another method uses two differently prepared states, the state of interest and a second state that is the result of operating on the original state.  Correlations can then be witnessed by comparing the two with distance measures \cite{LPB2010,SBPV2010,GB2011,D2011,GB2013,WLB2013}. This method does not make any assumptions about the state of the environment nor the structure of the system-environment correlations and has been shown to be experimentally successful at detecting initial correlations  \cite{LTG2011,SBCV2011,GRPBBH2014}.

Yet another recently discovered method can be used to find correlations between a system and its environment by only measuring the system if one makes the experimentally reasonable assumptions that (i) the system-environment state can be repeatedly prepared, (ii) the system can be measured, and (iii) that the system-environment interaction is known. It was found that when the system and environment consist of one qubit each, there is a wide range of correlated initial states that can be detected \cite{chit.2}.

Here we develop this theory for practical use and examine the experimentally relevant Hamiltonian sometimes called the Heisenberg exchange interaction.  This Hamiltonian, which essentially has the form $\vec{S}_1\cdot\vec{S}_2$ is relevant for spin-spin interactions such as those one may find in a system where the electron spin is correlated with a nuclear spin.  We consider various initial states of the spin-1/2 system and spin-1/2 environment $\rho_{SE}$ and ask how correlations between the system and environment might be detected by letting the system and environment evolve under this Hamiltonian for some time $t$.  We then present an analysis of the robustness of our results showing that it is indeed possible to detect correlations by measuring only the system.


\section{Detecting Initial Correlations}
\label{sec:detecting}

Our general method is described by the following.  We assume that the combined system and environment state, $\rho_{SE}$, evolves unitarily.  The evolution of $\rho_S=tr_E(\rho_{SE})$ can then be described by 
\begin{equation}
\label{eq:transformation}
    \rho_S(t_1) \rightarrow {\rho_S}'(t) = tr_E[U_{SE}(\rho_{SE}){U^\dagger_{SE}}].
\end{equation}
If the combined system that evolves as described above is initially uncorrelated, then 
\begin{equation}\label{eq:prodst}
    \rho_{SE} = \rho_S \otimes\rho_E.
\end{equation}


In either case we describe the initial state of the system by 
\begin{equation}
\rho_S = \tr_E(\rho_{SE}),
\end{equation}
and the final state of the system by
\begin{equation}
\rho_S^\prime = \tr_E(U_{SE}\rho_{SE}U_{SE}^\dagger).
\end{equation}
Furthermore, we assume that each of these can be measured using standard tomographic techniques.  The presence of initial correlations between the system and environment in $\rho_{SE}$ means that this combined state is not in a product state $\rho_{SE}\neq \rho_S\otimes\rho_E$.  Thus, given an initial state $\rho_S$ = $tr_E[\rho_{SE}]$ and the final state $\rho'_S$ as described in Eq.~(\ref{eq:transformation}), 
the presence of initial correlations between $\rho_S$ and $\rho_E$ can be determined by checking if the same $\rho'_S$ can be obtained by considering the transformation of an uncorrelated state  $\Tilde{\rho}_{SE}=\rho_S\otimes\Tilde{\rho}_E$.  In our examples below, we take the system to be one qubit and the environment to be an unknown two-state system.  The hypothetical arbitrary environment state, $\Tilde{\rho}_E$, is written as  
\begin{equation}\label{eq:arbenvironment}
\Tilde{\rho}_E=\frac{1}{2}(\textbf{I}+x\textbf{X}+y\textbf{Y}+z\textbf{Z}),
\end{equation}
where \textbf{X}, \textbf{Y}, and \textbf{Z} correspond to the Pauli operators and $x,y,z$ are real numbers.  In other words, we want to know if it is possible to find a $\Tilde{\rho}_E$ such that Eq.~(\ref{eq:transformation}) is satisfied.  Since we make no assumptions about the state of the environment, if we are able to obtain the same $\rho'_S$ with this uncorrelated $\Tilde{\rho}_{SE}$, then it is possible that our system-environment state of interest, $\rho_{SE}$, was uncorrelated all along. Conversely, if we are not able to produce $\rho'_S$ for any realistic $\Tilde{\rho}_E$, then the combined state is shown to possess some correlations between the system and environment.  

We are concerned here not necessarily with how correlations between the system and environment have arisen; it can be presumed that some previous interaction has produced these correlations. However, for the purposes of this paper we choose various $\rho_{SE}$ whose system-environment correlations are rather experimentally realistic. Each of these could describe a different experimentally prepared initial state that had some initial strong coupling between the system and environment or was not accurately prepared.

Consider two two-state systems interacting via the Heisenberg exchange Hamiltonian with some coupling constant $J$ which determines the interaction strength:
\beq
H_{ex} = J(\textbf{XX} + \textbf{YY} + \textbf{ZZ}),
\eeq
which gives rise to the dipole-dipole interaction and is of particular interest because of the pervasiveness of such interactions in experiment. The time evolution of the state corresponds to $U$ which, for simplification, we express in terms of a parameter $\alpha = Jt$, with $J$ the coupling constant and $t$ time:
\begin{equation}\label{DDHUnitary}
U(\alpha) = \begin{pmatrix}
e^{-i\alpha} &  & 0 & 0\\
0 & e^{i\alpha}\cos{2\alpha} & {-i}e^{i\alpha}\sin{2\alpha} & 0 \\
0 & {-i}e^{i\alpha}\sin{2\alpha} & e^{i\alpha}\cos{2\alpha} & 0 \\
0 &  & 0 & e^{-i\alpha}\\
\end{pmatrix}. 
\end{equation}
Note that $U(\pi/4)$ is the SWAP operator (times an overall phase $e^{i\pi/4}$ which is irrelevant).  Since nothing is assumed about the state of the environment, we can simulate any evolution when the SWAP operator acts on initial product state of the form $\Tilde{\rho}_E = \rho^\prime_S$.  In other words, if we suppose the final state of $S$ is $\rho_S^\prime$, then the initial state of the system plus environment $\rho_{SE}$ can be taken to be $\rho_S\otimes \rho_S^\prime$.  Also, $U(\pi/2)=i\bf{I}$.  So, this evolution can also always be simulated with a product state since $\rho_S^\prime=\rho_S$.

In this work we will show how to detect initial correlations between the system and environment undergoing this unitary transformation for three different states $\rho_{SE}$. For each $\rho_{SE}$, we will compare the state $\Tilde{\rho}'_S\equiv \tr_E(U\rho_S\otimes \tilde{\rho}_EU^\dagger)$, that is produced by the transformation on the uncorrelated $\Tilde{\rho}_{SE}$, to the $\rho_S^\prime$ that is the result of the transformation of $\rho_{SE}$, which may or may not be correlated. The difference between these two states is $D \equiv \rho'_S-\Tilde{\rho}'_S$ corresponding to each $\rho_{SE}$.  If $D=\textbf{0}$, we are not able to detect initial correlations between the system $\rho_S$ and its environment in $\rho_{SE}$.  

Let us emphasize that the difference is between the output of the experiment, $\rho'_S$, and the possible states $\tilde{\rho}'_S$, arising from $\Tilde{\rho}_{SE}$ account for all $\Tilde{\rho}_E$ making up a possible initial product state for the combined system and environment. Thus $D$ can be used to construct a distance $d$ between the state $\rho'_S$ and the set of possible final states generated by $\tr_E(U\Tilde{\rho}_{SE}U^\dagger)$: 
$$
d\equiv\underset{n^2\leq 1}{\mbox{min}}\left[\sum_{i}|D_{ij}|^2\right]^{1/2},
$$ 
where $n^2 = x^2+y^2+z^2$.


\subsection{Maximally entangled $\rho_{SE}$}
\label{sec:maxent}

For the first and motivational example of how initial correlations can be detected, we consider
\begin{equation}\label{state1}
    \rho_{SE} = \ket{\Psi}\bra{\Psi},
\end{equation}
where $\ket{\Psi} = \frac{1}{\sqrt{2}}(\ket{01}+i\ket{10})$. $\ket{\Psi}$ is a maximally entangled state that is locally equivalent to the Bell state $\ket{\Psi^+}$. Clearly, $\rho_{SE}$ contains correlations (since maximal entanglement produces the strongest correlations) between the system and environment.

Here $\rho_S = \textbf{I}/2$ and we suppose the state $\rho_{SE}$ evolves according to Eq. (\ref{eq:transformation}).  The final state is $\rho'_{S} = \frac{1}{2}(1-\sin{(4\alpha)})\ket{0}\bra{0}+\frac{1}{2}(1+\sin{(4\alpha)})\ket{1}\bra{1}$. In matrix form,
\begin{equation}\label{1rhoSprime}
\rho'_S = \begin{pmatrix}
\frac{1}{2}(1-\sin{(4\alpha)}) & 0\\
0 & \frac{1}{2}(1+\sin{(4\alpha)})\\
\end{pmatrix}. 
\end{equation}
Choosing $\alpha$ will determine $\rho'_S$ - we can easily see that $\rho'_S=\ket{0}\bra{0}$ when $\alpha=\frac{3\pi}{8}$.
In order to detect initial correlations (which we know to be present in a maximally entangled state), we must see if an uncorrelated $\Tilde{\rho}_{SE}$ can create the same evolution. $D$ comparing this state $\Tilde{\rho}'_S$ and the original $\rho'_S$ is
\begin{equation}\label{1D_eff}
D = \frac{1}{4}\begin{pmatrix}
A_{00} & A_-\\
A_+ & -A_{00}\\
\end{pmatrix},
\end{equation}
where $A_{00}=z(-1+\cos{(4\alpha)})+2\sin{(4\alpha)}$ and $A_{\pm}=-2(x{\pm}iy)\sin^2{(2\alpha)}$. Note that $D$ is sometimes, but not always, equal to \textbf{0}, depending on the values of $\alpha$, $x$, $y$, and $z$.

Recall that $x,y,z$ are parameters that determine the arbitrary environment state in Eq. (\ref{eq:arbenvironment}). If, for a value of $\alpha$, $x,y,z$ can be chosen such that $\Tilde{\rho}_{SE}$ can model the transformation and $\rho'_S$ is obtained, this will give $D=\textbf{0}$.  
It is thus evident that for us to be able the detect initial correlations in $\rho_{SE}$ for a given value of $\alpha$, there must not be any choice of $x,y,z$ such that $x^2+y^2+z^2\leq 1$ and $D=\textbf{0}$.  If there is such a set of $x$, $y$ and $z$ that will allow the uncorrelated $\Tilde{\rho}_{SE}$ to undergo the same transformation $\rho_S\rightarrow\rho'_S$, then we cannot state that there were initial correlations. If we cannot find such $x$, $y$, $z$, we know that there are initial correlations in $\rho_{SE}$.  This means we can distinguish between $\rho'_S$ and $\Tilde{\rho}'_{S}$ (i.e. $D$ will not be \textbf{0}).  Therefore, the task is to try to find $\Tilde{\rho}'_{S}$ with a valid $\Tilde{\rho}_E$ such that $D=\textbf{0}$.  For example, if $\alpha=\frac{3\pi}{8}$ then $\rho'_S$ is a pure state (as noted previously) while $\Tilde{\rho}'_{S}$ remains mixed. Recall that $\alpha$ depends on both the coupling constant and the time over which the interaction occurs; these two parameters may be ascertained or varied to obtain a particular $\alpha$.  

For this example, if $\alpha = \frac{3\pi}{8}$, then $D$ is 
\begin{equation}\label{1D_effz3pi8}
D = \begin{pmatrix}
\frac{1}{4}(-2-z) & \frac{1}{4}(-x+iy)\\
\frac{1}{4}(-x-iy) & \frac{1}{4}(2+z)\\
\end{pmatrix}. 
\end{equation}
The only possible combination of $x$, $y$ and $z$ for which $D=\textbf{0}$ is $x=y=0$, $z=-2$, but this solution does not satisfy the necessary condition that $x^2+y^2+z^2\leq1$ ($\Tilde{\rho}_E$ must be a valid density matrix), therefore we know that the unitary is effective in detecting initial correlations for $\alpha=\frac{3\pi}{8}$.

\begin{figure}[h]
    \centering
    \includegraphics[width=7.5cm]{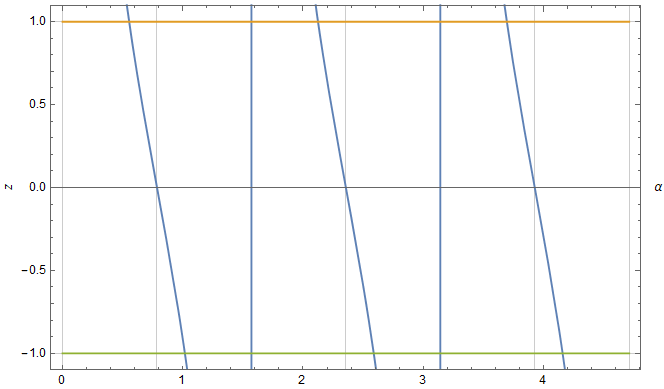}
    \caption{The blue function shows the relationship between $\alpha$ and the solution for $z$ ($x=y=0$) at which $D=\textbf{0}$ for $\rho_{SE}$ as in Eq. (\ref{state1}). The orange and green lines represent the restriction on $z$ because $x^2+y^2+z^2\leq1$. Note the periodic behavior.}
    \label{fig:1z-valueneeded}
\end{figure}

Varying $\alpha$ changes the evolution undertaken by the state and we can find which values of $\alpha$ will not enable us to detect initial correlations.  
By considering all possible values of $\alpha$ and the associated solution for $x,y,z$ such that $D$ (which is $\frac{\pi}{2}$-periodic) equals \textbf{0}, we can determine the efficacy of our method for those values. Consideration of $D$ in Eq.~(\ref{1D_eff}) shows that the solution will always require $x=y=0$ unless $\sin(2\alpha)=0$ (i.e. $U=i\textbf{I}$) and thus, for those $\alpha$ such that $\sin(2\alpha)\neq 0$, we only need to consider $z$. The relationship between $z$ and $\alpha$ can be seen graphically in Fig. \ref{fig:1z-valueneeded}.  

Numerically we can say that, if $D=\textbf{0}$, then
\begin{equation}\label{eq:1zalpha}
    z=\frac{-2\sin{4\alpha}}{\cos{4\alpha}-1}.
\end{equation}{}
By considering the restriction on $z$ i.e. $-1\leq{z}\leq1$ (because $x=y=0$ and $x^2+y^2+z^2\leq1$), we can calculate bounds on $\alpha$ for which $D\neq\textbf{0}$: 
\begin{equation}\label{1alphaboundsyes}
    \frac{1}{2}\arctan(-2)+\frac{n\pi}{2}\leq\alpha\leq\frac{1}{2}\arctan(2)+\frac{n\pi}{2},
\end{equation}
for any positive or negative integer $n$, excepting $\alpha=\frac{\pi}{2}\pm\frac{n\pi}{2}$, for which $D$ is always $\textbf{0}$. As expected, $\alpha=\frac{3\pi}{8}$ lies within the range of acceptable values. 


\subsection{Pure state mixed with a maximally entangled state}\label{sec:puremix}

\begin{figure}[h!]
    \centering
    \includegraphics[width=7.5cm]{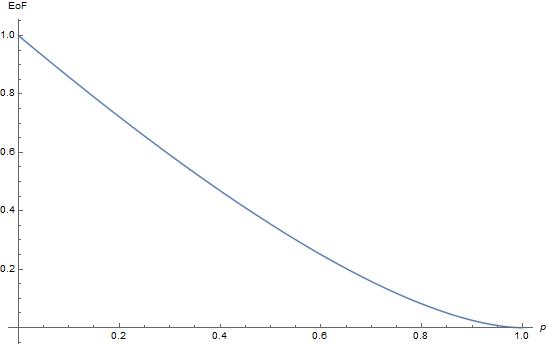}
    \caption{The entanglement of formation (EoF) contained in the state $\rho_{SE}$ (Eq. (\ref{state2})) as a function of $p$. When $p=0$, the maximum amount of entanglement is present.}
    \label{fig:2EofF}
\end{figure}

We now consider a mixture of a pure state with a maximally entangled state.  This $\rho_{SE}$ is of particular interest because of its connection to an experiment in which there is an attempt to create a pure state, but there are some interactions between the system and environment that produce correlations.  Here, too, we will show that evolution by the same unitary $U$ (Eq.~(\ref{DDHUnitary})) will result in detectable correlations.  Thus, we take 
\begin{equation}\label{state2}
    \rho_{SE} = p \ket{01}\bra{01} +\frac{(1-p)}{2}(\ket{01}+i\ket{10})(\bra{01}-i\bra{10}).
\end{equation}
Here, $\rho_{SE}$ may or may not be correlated depending on the value of $p$, where $0\leq{p}\leq{1}$. If we choose $p=0$, we obtain the entangled state in Eq. (\ref{state1}). The entanglement of this state is dependent on $p$ and can be calculated as described in Appendix~\ref{sec:appendix} and is graphically depicted in Fig. \ref{fig:2EofF}.

One can easily determine that
\begin{equation}\label{state2init}
\rho_S=p\ket{0}\bra{0}+\frac{1-p}{2}(\ket{0}\bra{0}+\ket{1}\bra{1})   
\end{equation}
and
\begin{equation}\label{state2final}
    \rho'_S = \frac{1}{2}\begin{pmatrix}
\rho_+ & 0\\
0 & \rho_-\\
\end{pmatrix},
\end{equation}
where $\rho_{\pm}=1{\pm}p\cos{(4\alpha)}\pm(-1+p)\sin{(4\alpha)}$. $D$ between this state and an initially uncorrelated state obtained by the evolution of $\Tilde{\rho}_{SE}=\rho_S\otimes\Tilde{\rho}_E$ is now
\begin{equation}\label{Deff2}
D = \frac{1}{2}\begin{pmatrix}
-B_{00} & B_-\\
B_+ & B_{00}\\
\end{pmatrix},
\end{equation}
where $B_{00}=(p+z){S_2}^2+2(p-1)S_2C_2$ and $B_\pm=2(x{\pm}iy)(-1+C_4{\pm}ipS_4)$. Here $S_i = \sin{(i\alpha)}$ and $C_i = \cos{(i\alpha)}$. It is clear that for this matrix to be \textbf{0}, $x$ and $y$ will once again be $0$, and the value of $z$ depends on both $\alpha$ (as before) and $p$.

For example, if $\alpha=\frac{3\pi}{8}$, a value that enabled the detection of initial correlations for $\rho_{SE}$ in Eq.~(\ref{state1}),
\begin{equation}\label{Deff23pi/8}
D = \frac{1}{4}\begin{pmatrix}
(-2+p-z) & (ix+y)(i+p)\\
(-ix+y)(-i+p) & (2-p+z)\\
\end{pmatrix}.
\end{equation}
For this value of $\alpha$, it can be easily seen that $D=\textbf{0}$ when $x=y=0$ and $z=p-2$ and, because $p$ ranges between 0 and 1, $-2\leq{z}\leq{-1}$. $z$ depends linearly on $p$. However, $z=-1$ is the only physically possible value for $z$ in this range and thus $D$ can only be made \textbf{0} when $p=1$, which represents a completely uncorrelated $\rho_{SE}$ (see Eq. (\ref{state2})). Therefore, this value of $\alpha$ is effective for the detection of initial correlations present in the initial state for a $\rho_{SE}$ of this form.

However, it is ultimately the measurement of the initial and final system states that indicates the presence of initial correlations to the experimenter, therefore $\rho_S$ and $\Tilde{\rho}_S$ must be distinguishable. When $\alpha=\frac{3\pi}{8}$ is plugged is used in Eq. (\ref{state2final}), we find that
\begin{equation}\label{state2final3pi8}
\rho'_S = \frac{1}{2}\begin{pmatrix}
2-p & 0\\
0 & p\\
\end{pmatrix}.
\end{equation}
This in turn shows that large correlations ($p\approx0$) are easily distinguishable - as Eq. (\ref{state2final3pi8}) will be close to a pure state - while small correlations will be more difficult to detect: the mixed state obtained when $p\approx1$ is close to the maximally mixed state that presents the possibility of no correlations.

\begin{figure}[h]
    \centering
    \includegraphics[width=7.5cm]{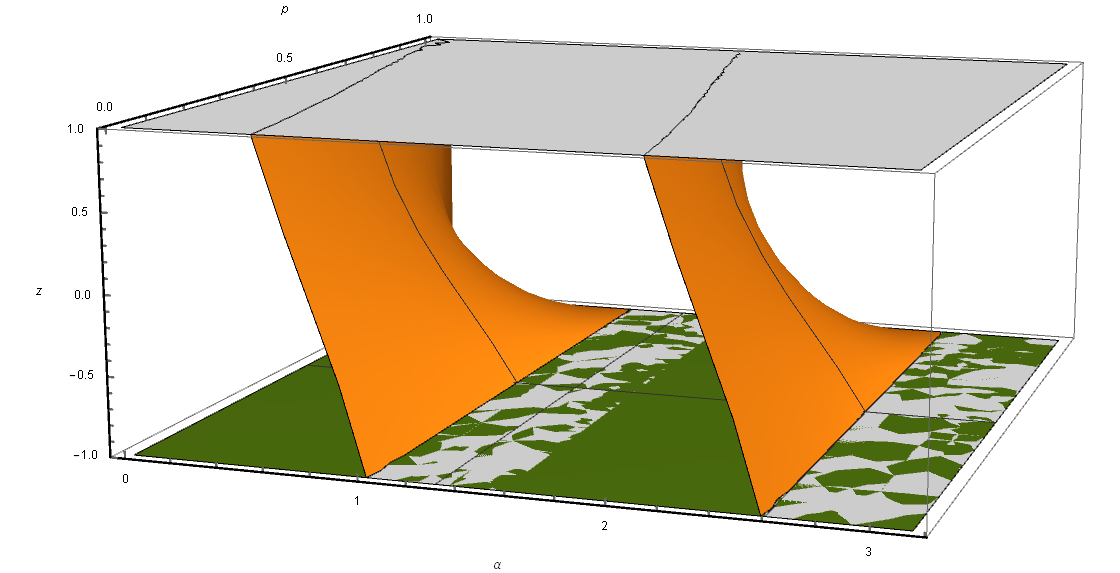}
    \caption{Relationship between $p$, $\alpha$, and $z$ (vertical axis) within between $z=-1$ and $z=1$ in the range of $0\leq\alpha\leq\pi$ when $x=y=0$. Note the discontinuity at $p=1$.}
    \label{2pandalphaz}
\end{figure}

Generally (i.e. for all values of $\alpha$), the dependence of $z$ on $\alpha$ and $p$ for which $D=\textbf{0}$ can be described by the following (a simplification of the diagonal terms of Equation (\ref{Deff2})):
\begin{equation}\label{2pazrelation}
z=-2(p-1)T_2-p = 2T_2-p(2T_2+1),
\end{equation}
where $T_2=\cot{(2\alpha)}$. This relationship is represented graphically in Fig. \ref{2pandalphaz}. Fig. \ref{fig:1z-valueneeded} is easily derived from the graph in Fig. \ref{2pandalphaz} when $p=0$. 

We once again find the range of values of $\alpha$ for which all correlations between the system and environment of this state can be detected, that is to say, no values of $x,y,z$ can be found such that $D=\textbf{0}$ unless $p=1$:
\begin{equation}\label{2newalphaboundsyes}
  \frac{\arccot(\frac{-1}{2})}{2}+\frac{n\pi}{2}\leq\alpha\leq\frac{\arccot(\frac{p+1}{1-p})}{2}+\frac{n\pi}{2},    
\end{equation}{}
for any positive or negative integer $n$ and excepting (see the discussion concerning $U$ at $\alpha=\frac{\pi}{4},\frac{\pi}{2}$)
$$
\alpha = \frac{\pi}{4}\pm\frac{n\pi}{4}.
$$

The behavior at $p=1$ is best determined by reasoning opposite to the analysis of parameter $z$ in the discussion surrounding Eq.~(\ref{Deff2}): The uncorrelated state is always modeled by the product state with $z=-1$.  

Our previously analyzed choice of $\alpha=\frac{3\pi}{8}$ works perfectly fine at detecting all initial correlations, and since for this state, entanglement is presented for $0\leq{p}<1$ (see Fig.~\ref{fig:2EofF}), the evolution caused by $U$ also indicates the presence of entanglement here. Of course, the bounds imply that there are other $\alpha$ which indicate initial correlations. However, values that lie closer to $\alpha=\frac{\pi}{2}$ for which $D$ is always $\textbf{0}$ (see the restriction on Eq.~(\ref{2newalphaboundsyes})), may be less favorable due to the potential for error.


\subsection{Maximally mixed state}

\begin{figure}
    \centering
    \includegraphics[width=7.5cm]{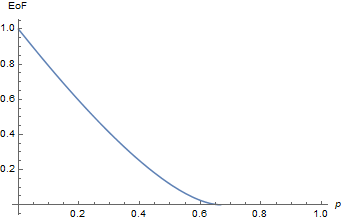}
    \caption{The graph above shows the entanglement of formation vs. p for the state in Equation (\ref{state3}).}
    \label{fig:3eoff}
\end{figure}

We now turn our attention to performing the protocol with the following initial state,
\begin{equation}\label{state3}
    \rho_{SE}=p\frac{1}{4}\textbf{I} +\frac{1-p}{2}(\ket{01}+i\ket{10})(\bra{01}-i\bra{10}).
\end{equation}
The amount of entanglement present in this state is once again dependent on $p$, as can be seen in Fig. \ref{fig:3eoff}, and is $0$ for $p\geq{2/3}$. It would therefore be useful for our protocol to be able to detect correlations within the range of $p$ where entanglement is present, but also beyond this range, all the way to (but excepting) $p=1$, as the state is correlated for all values $0\leq{p}<1$.

{}

\begin{figure}[h]
    \centering
    \includegraphics[width=7.5cm]{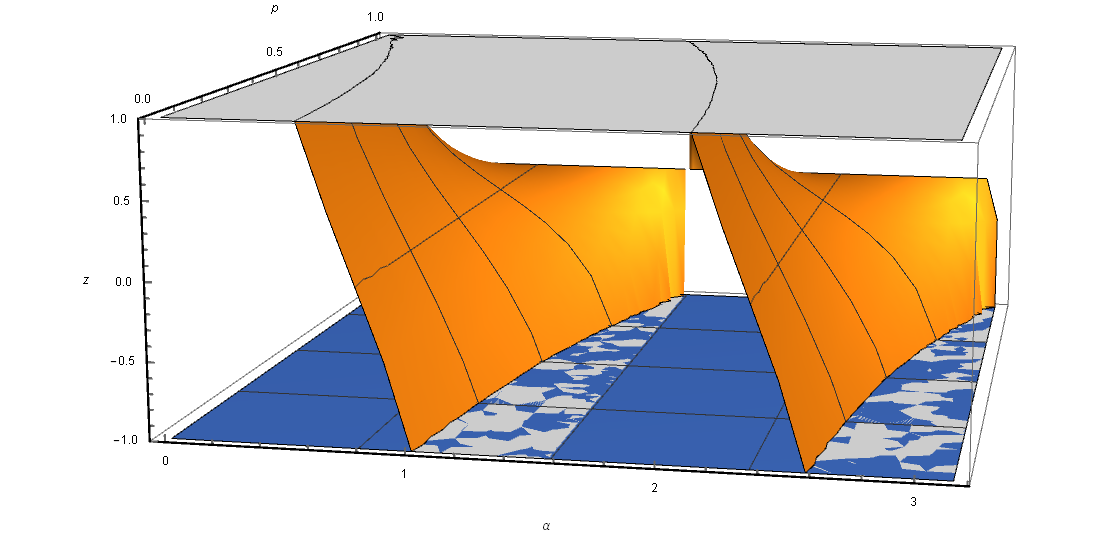}
    \caption{Relationship between parameters $z$, $p$ and $\alpha$ for $D$ corresponding to the state in Eq. (\ref{state3}) for $x=y=0$ in range $0\leq\alpha\leq\pi$. Note behavior at $\alpha=\frac{\pi}{2}$ ($U=i\textbf{I}$).}
    \label{fig:state3_p_a_z}
\end{figure}

Conducting a similar analysis as with the previous two examples, the final state is  
\begin{equation}\label{3rhoS'}
    \rho'_S = \frac{1}{2}\begin{pmatrix}
1+(-1+p)S_4 & 0\\
0 & 1-(-1+p)S_4\\
\end{pmatrix},    
\end{equation}{}
where $S_4=\sin{(4\alpha)}$ once again, and
\begin{equation}\label{3D_eff}
D=\frac{1}{2}\begin{pmatrix}
-C_{00} & -(x-iy){S_2}^2\\
-(x+iy){S_2}^2 & C_{00}\\
\end{pmatrix},    
\end{equation}
where $C_{00}=S_2(2(-1+p)C_2+zS_2)$, $S_2=\sin{(2\alpha)}$, and $C_2=\cos{(2\alpha)}$. The off-diagonal of Eq. (\ref{3D_eff}) implies that all solutions for $D=\textbf{0}$ must have $x=y=0$. The diagonal terms indicate the relation between $z$, $p$, and $\alpha$:
\begin{equation}\label{3zpalpha}
    z = -2(-1+p)\cot(2\alpha),
\end{equation}
which is the very similar to the corresponding relation in Part \ref{sec:puremix}, Eq.~(\ref{state2}) (the latter includes an additional component). Eq.~(\ref{3zpalpha}) is graphically represented by Fig.~\ref{fig:state3_p_a_z}.

We are once again tasked with finding a value of $\alpha$ that for which we are unable to find a $z$ satisfying $D=\textbf{0}$, indicating the ability to distinguish between an uncorrelated and correlated $\rho_{SE}$.

The value of $\alpha$ that was successful in doing so for the states in Parts \ref{sec:maxent} and \ref{sec:puremix}, $\frac{3\pi}{8}$, is no longer successful at detecting correlations here for all values of $p$. This is evident upon considering $D$ when $\alpha=\frac{3\pi}{8}$:
\begin{equation}\label{3Deff3pi8}
        D = \begin{pmatrix}
\frac{1}{4}(-2+2p-z) & \frac{1}{4}(-x+iy)\\
\frac{1}{4}(-x-iy) & \frac{1}{4}(2-2p+z)\\
\end{pmatrix}.  
\end{equation}
When $p\geq 0.5$,  i.e. only weak correlations are present, a $z$ value can be found which models the transformation with an uncorrelated initial state ($D=\textbf{0}$, and thus the protocol fails at detecting the correlations in states where $p\geq0.5$). Can this restriction on $p$ for which correlations are detected be relieved?

By considering the $p$ and $z$ relations for various $\alpha$, we verify their periodicity and at the same time find that our ability to detect correlations increases as $\alpha$ approaches, but does not reach, $\frac{\pi}{2}$. For example, when $\alpha = \frac{13\pi}{32}$, the limitation on $p$ becomes $p\leq\frac{2}{3}$, and for $\alpha =  \frac{15\pi}{32}$, it is approximately $p\leq.90$ for the protocol to detect the correlations.

Approaching $\alpha=\frac{\pi}{2}$ even more, we have $p<.95075$ when $\alpha=\frac{31\pi}{64}$, and $p<.9754$ for $\alpha=\frac{63\pi}{128}$. These latter values would, in theory, allow us to detect even the smallest system-environment correlations present in $\rho_{SE}$.

Realistically detecting slight correlations, however, is difficult as when $\alpha=\frac{\pi}{2}$, no correlations can be detected. Therefore, the protocol for detecting most to all correlations for this state is not very fault-tolerant - choosing $\alpha$ close to but not equal to $\frac{\pi}{2}$ is difficult to achieve in experiment  without error.

Interesting, however, is that for $\alpha=\frac{13\pi}{32}\pm\frac{n\pi}{2}$ (for any positive or negative integer $n$), the range for which correlations can be detected is most nearly $p<\frac{2}{3}$. As mentioned previously and depicted in Fig. \ref{fig:3eoff}, this is also the bound on $p$ for which entanglement is present. Performing the protocol with $\alpha=\frac{13\pi}{32}$ could therefore be used as a way to detect entanglement in the state. This choice of $\alpha$ is also more fault-tolerant - a small error in this value, likely to occur in an experiment, would not make it equal to $\frac{\pi}{2}$, for which $D$ is always $\textbf{0}$ and therefore no correlations at all would be detected.  


\section{Robustness of Results}

In order to examine the robustness of the results described in the previous section, we follow the technique described in \cite{OpenSysRobust}. We let an uncorrelated state $\tau_S\otimes\rho_E$ evolve unitarily according to Eq.~(\ref{eq:transformation}) and we obtain the final state $\tau'_S$. $\tau_S$ and $\rho_E$ are both arbitrary states, where 
\begin{equation}\label{tauS}
    \tau_S = \frac{1}{2}(\textbf{I}+{\vec{n}\cdot\vec{\sigma}})
\end{equation}
and
\begin{equation}\label{rhoE}
    \rho_E = \frac{1}{2}(\textbf{I}+{\vec{m}\cdot\vec{\sigma}}).
\end{equation}
After the state has evolved, we can compare the final and initial states $\tau_S$ and $\tau'_S$ to their respective counterparts, $\rho_S$ and $\rho'_S$, by varying values for $n_1,n_2$ and $n_3$ ($m_1,m_2,m_3$). If we cannot equate both pairs using one set of values for $n_i$ and $m_i$, the example is robust to errors.

Evolving $\tau_S\otimes\rho_E$ with the unitary in Eq. (\ref{DDHUnitary}) and tracing out the environment gives us 
\begin{equation}\label{tauSprime}
    \tau'_S = \frac{1}{4} \begin{pmatrix}
\tau'_{S00+} & \tau'_{S01-}\\
\tau'_{S01+} & \tau'_{S00-}\\
\end{pmatrix},
\end{equation}
where 
\begin{eqnarray}
\tau'_{S00\pm}&=&2{\pm}(m_3+n_3+(-m_3+n_3)\cos{(4\alpha)}\nonumber \\
&&+(-m_2n_1+m_1n_2)\sin{(4\alpha)})\nonumber
\end{eqnarray}
and 
\begin{eqnarray}
\tau'_{S00\pm}&=&m_1{\pm}im_2+n_1{\pm}in_2\nonumber \\
&&+(-m_1{\mp}im_2+n_1{\pm}in_2)\cos{(4\alpha)}\nonumber \\
&&+(im_3n_1{\mp}m_3n_2-im_1n_3{\pm}m_2n_3)\sin{(4\alpha)}).\nonumber
\end{eqnarray}

Let us consider the first example, i.e. Eq. (\ref{state1}). Comparing the initial state $\rho_S = \frac{\textbf{I}}{2}$ to Eq. (\ref{tauS}), we find that to equate the two, $|\vec{n}|$ must be small. Then, comparing Eq. (\ref{tauSprime}) to Eq. (\ref{1rhoSprime}), we find that $m_2n_1=1$. If $|\vec{n}|$ is small, we can take $n_1=\epsilon$, some small value. Then, $m_2=\frac{1}{\epsilon}$, which is greater than 1. Since $|\vec{m}|\leq1$, this is not an acceptable value for $m_2$ and the result is robust. There is another possible approach towards a solution here, which produces the same result. 

By equating Eq. (\ref{state2init}) and Eq. (\ref{tauS}) (the second example), we find that $n_3=p$, while $n_1,n_2$ approach $0$ and thus can be considered to have some small value $\epsilon$ as long as $|\vec{n}|\leq1$ holds true. From our comparison of $\rho'_S$ and $\tau'_S$ (Eqs. (\ref{state2final}) and (\ref{tauSprime})), we see that $m_3=-p$. Finally, to completely equate the two final states, we must satisfy the following condition:
\begin{equation}\label{condition2}
\frac{-m_2n_1+m_1n_2}{2}=-1+p.    
\end{equation}
Since both $n_1$ and $n_2$ approach $0$ and we have already seen that neither $m_2$ nor $m_1$ are able to have a value of $\frac{1}{\epsilon}$ for some small value $\epsilon$ in the first example, this is satisfied only when $p=1$. This value of $p$ is the completely uncorrelated state, and thus the only value for which we expect to be able to equate $\rho_S$ and $\tau_S$. This example as well is robust.

Finally, for the initial third state, we must equate the partial trace over the environment of Eq.~(\ref{state3}) with Eq.~(\ref{tauS}). Once again, we derive that $|\vec{n}|$ must be small. As for the final state, so that Eq. (\ref{3rhoS'}) equals Eq. (\ref{tauSprime}), we know that $m_3$ must also be small. Then we are left to satisfy the same condition as in Eq. (\ref{condition2}). Even though our value for $m_3$ is different, the same reasoning as in the previous case applies and we are not able to find $m_1,m_2$ such that $|\vec{m}|\leq1$ unless $p=1$, the uncorrelated case, proving that the final example is also robust to errors in the initial and final states.


\section{Conclusion}

We have described a method which enables the detection of correlations between a system and environment using measurements only on the system.  This was shown to be effective for systems which evolve under a Heisenberg exchange Hamiltonian (Eq.~(\ref{eq:transformation})).  When a two-state system evolves from an initial state $\rho_S$ into a final state $\rho'_S$, correlations present between the system and environment can be detected.  This depends, of course, on the time that they evolve which we described by the parameter $\alpha$.  This may be exemplified experimentally by a dipole-dipole interaction. By varying $\alpha$, we can tune our protocol for detecting initial correlations.

We have have shown how our method detects correlations using three different initial states $\rho_{SE}$ (Eqs.~(\ref{state1}),(\ref{state2}),(\ref{state3})).  These different initial states describe models for a system interacting with its environment to produce correlations which we were able to vary using a parameter $p$.  For the $\rho_{SE}$ described in Eq.~(\ref{state1}) and in Eq.~(\ref{state2}), respectively, choosing $\alpha=\frac{3\pi}{8}$ enables us to state with certainty when correlations are present, although this is not the only value of $\alpha$ that enables this. For $\rho_{SE}$ as described in Eq. (\ref{state3}), choosing $\alpha$ to be close to but not equal to $\frac{\pi}{2}$ enables the detection of large, as well as small, correlations. Finally, a choice of $\alpha=\frac{13\pi}{32}$ for this state enables the detection of entanglement.  We have also shown that these results are robust to experimental error for properly chosen $\alpha$.  

Our work is different from early methods described in the introduction where initial correlations in an open system are detected without assumptions of its evolution but using multiple initial states. While we assume a particular form for the interaction undergone by the system, this is compatible with experimental reality as the experimenter is likely to have knowledge about the type of interactions the system undergoes, and, using this method, is able to detect correlations by considering only the state of interest.

Future work is required to describe the distinguishability of outcomes $\rho'_S$ for correlated and uncorrelated initial $\rho_{SE}$, as experimental measurement error can result in a lack of accuracy in the determination of the system's state. Moreover, this research can be expanded by looking further into the possibility of detecting entanglement and considering a broad selection of $\rho_{SE}$ that undergo this transformation. As discussed previously, this method can be used for other types of interactions that define the state evolution, although we appreciate our choice of unitary because of its connection to experiment.


\section{Acknowledgments}

Funding for this research was provided by the NSF, MPS under award number PHYS-1820870.  The authors thank Alvin Gonzales, Daniel Dilley, and Purva Thakre for many helpful discussions.


%


\appendix

\section{Measuring entanglement}\label{sec:appendix}

There are various so-called entanglement measures that can be used to quantify the amount of entanglement present in state $\rho$. One of these is the entanglement of formation $E$ (EoF), which is closely related to the concurrence $C$ of a pure state $\Phi$ as follows:

\begin{equation}
    E(\Phi)=\mathcal{E}(C(\Phi)),
\end{equation}

where $\mathcal{E}$ is defined by

\begin{equation}
    \mathcal{E}(C)=h\left(\frac{1+\sqrt{1-C^2}}{2}\right);
\end{equation}

\begin{equation}
    h(a)=-a\log_2a-(1-a)\log_2(1-a).
\end{equation}

The more general case involves a mixed state $\rho$, where $\mathcal{E}={E(\rho)}$ and $C(\rho)= max\{0,\lambda_1-\lambda_2-\lambda_3-\lambda_4\}$ and $\{\lambda_i\}$ are the square roots of the eigenvalues of $\Bar{\rho}$, which is in turn defined by

\begin{equation}
    \Bar{\rho}=\rho\Tilde{\rho};
\end{equation}

\begin{equation}
    \Tilde{\rho}=(\sigma_y\otimes\sigma_y)\rho*(\sigma_y\otimes\sigma_y),
\end{equation}

and * indicates taking the complex conjugate in the standard basis \cite{Wootters}.

\end{document}